\documentclass[aps,prl,twocolumn]{revtex4-1}

\usepackage{natbib}
\usepackage{graphicx}
\usepackage{amssymb, amsmath}
\usepackage{todonotes}
\usepackage{chapterbib}

\usepackage[normalem]{ulem}

\newcommand{\aap}{Astron.\ Astrophys.}
\newcommand{\mnras}{Mon.\ Not.\ R.\ Astron.\ Soc.}

\newcommand{\apjl}{Astrophys.\ J.\ Lett.}

\newcommand{\nphysa}{Nucl.\ Phys.\ A}

\definecolor{cerise}{rgb}{0.87, 0.19, 0.39}

\begin{document}

\title{Heat release in accreting neutron stars}

\author{M. E.\ Gusakov}
\affiliation{Ioffe Institute, St.-Petersburg, Russia}

\author{A. I.\ Chugunov}
\affiliation{Ioffe Institute, St.-Petersburg, Russia}

\date{\today}

\begin{abstract}
Observed thermal emission from accreting neutron stars (NSs) 
in a quiescent state is believed to be powered by nonequilibrium nuclear reactions 
that heat the stellar crust (deep crustal heating paradigm). 
We derive a simple universal formula for the heating efficiency, 
assuming that an NS has a fully accreted crust. 
We further show that, 
within the recently proposed thermodynamically consistent approach to the accreted crust, 
the heat release can be parametrized by the only one parameter 
--- the pressure $P_{\rm oi}$ at the outer-inner crust interface 
(as we argue, this pressure should not necessarily coincide with the neutron-drip pressure). 
We discuss possible values of $P_{\rm oi}$ 
for a selection of nuclear models that account for shell effects, 
and determine the net heat release and its distribution in the crust 
as a function of $P_{\rm oi}$. 
We conclude that the heat release should be reduced 
by a factor of few in comparison to previous works.
\end{abstract}


\maketitle

\textit{Introduction.--}
%
The crust of an accreting neutron star (NS) 
is driven out of thermodynamic equilibrium
by the accretion process from the NS companion. 
Its  composition is governed by exothermal nuclear reactions, 
which act to return 
it
back to the equilibrium.
It is generally believed that the heat released in these reactions  
is responsible for
the observed thermal luminosity of transiently accreting NSs 
\cite{hjwt07,gkm08,schatz_etal14,meisel_etal15,wdp17,mdkse18} (deep crustal heating paradigm \cite{bbr98}).
Observations of such stars could 
shed light on the properties of
superdense matter 
in their interiors \cite{ylpgc04,heinke_et_al_09,pr13,hp17,bbr98,pcc19}.

Physics of accreted crust (AC) is studied 
for about forty years since the pioneering works 
by Sato \cite{Sato79} and Haensel \& Zdunik \cite{HZ90,HZ90b}.
During this time
a number of 
AC models
were developed.
They are
based either on
detailed \cite{lau_ea18} 
or simplified \cite{Steiner12,SC19_MNRAS} reaction networks, 
theoretical atomic mass tables,
liquid-drop
approach
\cite{HZ03,HZ08},
or on the
density-functional theory 
\cite{Fantina_ea18}. 
The key question researchers want to answer:
What is the AC composition and heat  released in the crust
per accreted nucleon?
This question is usually addressed within the {\it traditional approach},
in which one follows the compositional changes in the given accreted fluid element
as it 
is compressed
under
the weight of 
newly accreted material. 
In this approach all constituents (nuclei, electrons, unbound neutrons) 
in the fluid element move together, with one and the same velocity.
Thus, traditional approach completely ignores the possibility that unbound neutrons, 
presented in the inner crust,
can redistribute 
themselves
independently of nuclei
in order to reduce the system energy.
In \cite{gc20} 
we reveal crucial importance of this effect
and show that neutrons in the inner crust 
must be in hydrostatic/diffusion equilibrium (nHD) state,
in which $\mu_n^\infty=\mu_n e^{\nu/2}=\mathrm{const}$,
where $\mu_n$ is the neutron chemical potential and $e^{\nu/2}$ is the redshift factor.
Equation of state (EOS) that respects the condition $\mu_n^\infty={\rm const}$ 
(hereafter nHD condition),
turns out to be
rather close to the 
catalyzed
EOS
of non-accreting (ground state) inner crust, 
and is very different from the traditional AC EOSs \cite{gc20}.

In this Letter we further explore the consequences of nHD
equilibrium.
First, we present a
universal formula (\ref{Q_dch}) 
showing that the 
deep crustal 
heat release $Q^\infty$,
as seen by a distant observer,
for a 
fully accreted crust (FAC)
is determined by 
EOS,
but not by
the details of nuclear transformations proceeding in the crust.
Second, we 
calculate 
the net heat release and 
its distribution over the nHD crust.

\textit{General energetics.--}
Consider an NS with AC and the mass $M$, 
and an NS with catalyzed crust, the mass $M_{\rm cat}$ 
and the same total amount of baryons, $A_{\rm b}$ 
(the cores of both NSs are assumed to be in the ground state).
The energy excess stored in the NS with AC is: 
$E^\mathrm{ex}=M-M_\mathrm{cat}$.
This energy will be released if one 
waits sufficiently long for the AC to relax to catalyzed crust 
by means of nonequilibrium nuclear processes.

To calculate the increase in $E^{\rm ex}$
associated with the accretion process, 
let us add 
$\delta A_b$ baryons (in the form of, e.g., H or He)
with the average mass per baryon $\overline{m}_{b}$.
This will increase $M$ by an amount:
$\delta M =\overline{m}_{b} \, \delta A_b \,e^{\nu_{ s}/2}$  \cite{ZN71,hpy07}, 
where $e^{\nu_{ s}/2}$ is the redshift factor
at the NS surface.
In turn, $M_\mathrm{cat}$ will increase, after adding the same $\delta A_b$ baryons, by 
$\delta M_\mathrm{cat} = \delta A_b\,\mu_{b,\mathrm{cat}}^\infty$  \cite{ZN71,hpy07}, 
where 
$\mu_{b,\mathrm{cat}}^\infty$
is the redshifted baryon chemical potential in the catalyzed NS, 
which 
is constant throughout the star \cite{ll80,ga06,gc20}. 
As a result, 
\begin{equation}
\delta E^\mathrm{ex}=(\overline{m}_b \,e^{\nu_{ s}/2}-\mu_{b,\mathrm{cat}}^\infty)\delta A_b.
\label{deltaEex} 
\end{equation} 
Generally, to determine what fraction of 
$\delta E^\mathrm{ex}$ 
goes into heat
and what is stored in the (nonequilibrium) crust, 
one should
study 
the kinetics of nuclear reactions there. 
However, in an important case of
an NS with FAC a general formula for the heat release can be derived. 

\textit{Universal heating formula for FAC}.--
In the process of accretion an NS eventually reaches the regime,
in which crust EOS does not further change noticeably in time.
We call such crust ``fully accreted''.
Subsequent accretion of material onto the surface of FAC
initiate nuclear reactions, which maintain the crust composition 
(see, e.g., \cite{gc20}).
The heat release associated with these reactions
can be calculated 
in analogy to the derivation of Eq.\ (\ref{deltaEex}).
Let us accrete $\delta A_b$ baryons onto the surface of an NS with FAC.
On the one hand, the energy of such star will change by
$\overline{m}_{b} \, \delta A_b \,e^{\nu_{ s}/2}$;
on the other hand, 
the star energy will vary by $\partial M/\partial A_b \, \delta A_b$,
where the partial derivative is taken, by assumption, 
at {\it fixed} EOS of FAC.
These two energies are not equal to one another; 
they differ by an amount of heat 
generated in the crust by nonequilibrium nuclear reactions, 
caused by accretion of $\delta A_b$ baryons.
Correspondingly, the total 
heat release per accreted baryon, redshifted to a distant observer,  is given by
\begin{equation}
Q_\mathrm{tot}^\infty= \overline{m}_b \,e^{\nu_{ s}/2}
- \partial M/\partial A_b.
\label{Q_acc}
\end{equation}
As shown in the Supplemental Material, 
$\partial M/\partial A_b$ can be presented as 
${\partial M}/{\partial A_b}=\mu^\infty_{b, {\rm core}}+\mathcal{O}(Q_{\rm tot}^\infty M_{c}/M)$, 
where $\mu^\infty_{b, {\rm core}}$ is the redshifted baryon chemical potential in the NS core;
$\mathcal{O}(Q_{\rm tot}^\infty M_{c}/M)$ 
is a small correction of the order of $Q_{\rm tot}^\infty M_{c}/M$; and $M_c$ is the crust mass.

In the upper layers of NSs (up to the density $\lesssim 10^9$~g\,cm$^{-3}$) 
the accreted material fuse into heavy nuclei 
(ashes with average mass per baryon 
$\overline{m}_{b,\mathrm{ash}}$) 
\cite{mdkse18}. 
The respective energy,  
$Q^\infty_\mathrm{ash}=(\overline{m}_b-
\overline{m}_{b,\mathrm{ash}})\,e^{\nu_{s}/2}$, 
is emitted from the surface without significant heating of the crust. 
The remaining part $Q^\infty \equiv Q_{\rm tot}^\infty-Q_{\rm ash}^\infty$ 
is released in the deep AC layers
\begin{eqnarray}
Q^\infty
&=&\overline{m}_{b,\mathrm{ash}}
\,e^{\nu_{s}/2}-
\partial M/\partial A_b
\nonumber\\
& =& \overline{m}_{b,\mathrm{ash}}\,e^{\nu_{s}/2}-
\mu_{b, {\rm core}}^\infty + \mathcal{O}(Q^\infty_{\rm tot} M_c/M).
\label{Q_dch}
\end{eqnarray}
It can be interpreted as an excess of gravitational energy, 
which ashes have at the outer layers of the accreted crust
plus some (typically small) excess of nuclear energy 
of the ashes with respect to the ground state composition ($^{56}$Fe).
Below,
we neglect
small corrections 
$\sim Q^\infty_{\rm tot} M_c/M \ll 1$~MeV in Eq.\ (\ref{Q_dch}),
but, in principle, they can be calculated.
Neglecting small energy carried away from the star in the form of neutrinos \cite{Gupta_ea07,HZ08,Fantina_ea18,SC19_MNRAS},
$Q^\infty$ represents the heat deposited in the crust.

Equation (\ref{Q_dch}) is
very general and can be applied to {\it any} FAC model.
In the Supplemental material we show that 
it 
reproduces,
in particular,
the results 
of
traditional one-component model \cite{HZ90,HZ90b,HZ03,HZ08,Fantina_ea18,zfh17}.
Equation (\ref{Q_dch})
can be further simplified for the nHD crust [see Eq.\ (\ref{Q_dch2_nHD})],
whose properties are briefly considered below.

\textit{The nHD crust: basic features.--}
The thermodynamically consistent model of the inner crust should 
respect
the nHD condition \cite{gc20}. 
To account for this condition, one should
substantially modify the traditional approach by
self-consistently analyzing 
nuclear processes in the {\it whole} inner crust, 
allowing for
redistribution of unbound neutrons over different crust layers and the core. 
Let us summarize the main
properties
of the nHD crust \cite{gc20}, 
which 
will be used in what follows.

The outer-inner crust interface (oi interface) plays an important role.
Above it unbound neutrons are absent 
and cannot travel between different layers, hence 
the traditional approach there is justified.
Below the oi interface unbound neutrons must redistribute in order to meet the nHD condition. 
At first glance the position of this interface 
should 
coincide with the point where neutrons drip out of nuclei in the traditional AC model.
However, as shown in \cite{gc20}, it is not the case:
unbound neutrons from the underlying layers can spread above this point
if it is energetically favorable.
Therefore, the position of the oi interface 
(parametrized by the pressure $P_\mathrm{oi}$)
should be considered as an additional parameter of the nHD crust model.
In particular, $P_\mathrm{oi}$ 
should vary over time
until an NS reaches the FAC state. 
In this state 
$P_\mathrm{oi}$ is fixed by the requirement that
the total amount of nuclei in the inner crust 
is almost constant during the accretion process
(otherwise EOS should evolve in time, 
which contradicts the FAC definition).
 
The process that keeps constant the number of nuclei in the crust
has been identified in \cite{gc20};
it is related to a specific instability,
which
disintegrate nuclei in the inner crust at the same rate as they are provided
by accretion onto the NS surface.

\textit{Thermodynamically consistent model 
of the inner crust: heat release.-- }
According to nHD condition
$\mu_{n}^\infty={\rm const}$ in the inner crust and core.
On the other hand, in the core $\mu_{n}=\mu_b$,
while at the oi interface (from the inner crust side) $\mu_n=m_n$.
In view of these facts, one has:
$\mu_{\rm b, {\rm core}}^\infty=m_n e^{\nu_\mathrm{oi}/2}$,
where $e^{\nu_\mathrm{oi}/2}$ is the redshift at the oi interface.
Now, expressing $\nu_s-\nu_\mathrm{oi}$ using one of the TOV equations \cite{hpy07},
Eq.\ (\ref{Q_dch}) 
can be rewritten as
\begin{equation}
Q^\infty=e^{\nu_\mathrm{oi}/2}
\left[
\overline{m}_{b,\mathrm{ash}}
\exp\left(\int_0^{P_\mathrm{oi}}\frac{dP}{P+\epsilon}\right)- m_n\right],
\label{Q_dch2_nHD}
\end{equation}
where $\epsilon=\epsilon(P)$ is the energy density and $P$ is the pressure.
In contrast to Eq.\ (\ref{Q_dch})
this formula
applies only to nHD crust.
It says that the heat release 
$Q^\infty$,
parametrized by the pressure  $P_\mathrm{oi}$,
can be easily found provided that
EOS in the outer crust (at $P<P_\mathrm{oi}$)
is known.
Note that the outer crust can be modeled within the traditional approach,
so that EOS there is relatively well established 
\cite{HZ08,Fantina_ea18,lau_ea18,Chamel_ea20_shallowHeating}.
Another form of the expression for
$Q^\infty$
is discussed in the Supplemental material, 
where we 
also present its independent microscopic
derivation 
valid for
the smoothed 
compressible liquid drop (CLD) model \cite{gc20}.

\begin{figure}
	\includegraphics[width=\columnwidth]{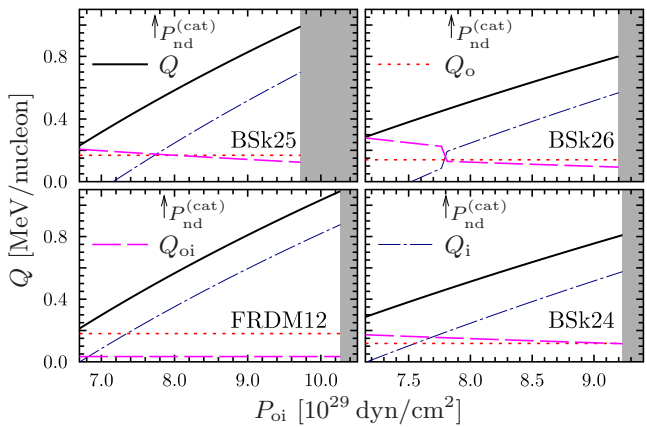}%
	\caption{
		The heat release vs $P_{\rm oi}$ for
		several nuclear models. 
		BSk24, BSk25, BSk26: 
		Hartree-Fock-Bogoliubov calculations \cite{HFB24};
		FRDM12:
		finite-range droplet macroscopic model \cite{FRDM12}.
		Arrows indicate the neutron drip pressure for catalyzed crust, $P_\mathrm{nd}^\mathrm{(cat)}$.
			For each model, the pressure region above the neutron drip point for AC
			(calculated
			within the
			traditional approach) is shaded gray.
		\label{Fig_Q}}
\end{figure}

To illustrate usefulness of the formula (\ref{Q_dch2_nHD}),
we 
calculate $Q^\infty$ for several nuclear models (BSk24, BSk25, BSk26: 
Hartree-Fock-Bogoliubov calculations \cite{HFB24};
FRDM12:
finite-range droplet macroscopic model \cite{FRDM12}).
The respective 
heat release
$Q$, defined as $Q=Q^\infty e^{-\nu_\mathrm{oi}/2}$,
is shown in Fig.\ \ref{Fig_Q} 
as a function of $P_\mathrm{oi}$
(for simplicity,
following \cite{HZ90,HZ08,Fantina_ea18}, pure $A=56$ composition 
of the ashes
is assumed).
For the region of $P_{\rm oi}$ depicted in the figure,
$Q\sim (0.2 - 1)$~MeV/nucleon,
being almost a linear function of $P_\mathrm{oi}$. It is
by a factor of few 
smaller than the heat release
$\sim (1.5-2)$~MeV/nucleon,
found in the traditional approach \cite{HZ08,lau_ea18,Fantina_ea18,SC19_JPCS}.
Note that the shell effects 
increase
the heat release
(for example, for the smoothed CLD model of \cite{gc20},
which ignores them,
we obtain
$Q\approx 0.11$~MeV/nucleon; see Supplemental material).
Similar feature was pointed out for the traditional AC model in \cite{Fantina_ea18}.

To analyze the heat release distribution in the FAC we do the following.
First, the net heat release in the outer crust ($Q_{\rm o}$) 
and its distribution can be easily found 
in
the traditional approach 
(see dots in Fig.\ \ref{Fig_Q}).
In turn, the heat release
at the oi interface ($Q_\mathrm{oi}$)
is associated with exothermic neutron absorptions
and electron emissions
by nuclei crossing the oi interface.
Neutrons, necessary for such absorptions
are supplied by continuous upward neutron flow in the inner crust.
The origin of the flow is the disintegration instability mentioned above. 
Neutrons, released in the course of this instability,
redistribute in the inner crust and core in order to maintain the nHD equilibrium (for details see \cite{gc20}).
To find $Q_\mathrm{oi}$,
we slightly modified the reaction network of \cite{SC19_MNRAS}
by allowing for absorptions of arbitrary number of neutrons 
at the oi interface by incoming nuclei 
(in the spirit of the extended Thomas-Fermi plus Strutinsky integral (ETFSI) calculations of \cite{Fantina_ea18}).
The resulting $Q_\mathrm{oi}$ is shown 
in Fig.\ \ref{Fig_Q} by 
long 
dashes.
The remaining heat, $Q_{\rm i}=Q-Q_{\rm o}-Q_\mathrm{oi}$,
is released in the inner crust and is shown by dot-dashed lines.

\textit{$P_\mathrm{oi}$ determination and heat release distribution in the inner crust.-- }
We demonstrate that $Q_{\rm o}$, $Q_{\rm i}$, and $Q_\mathrm{oi}$ 
are fully determined by 
the pressure $P_\mathrm{oi}$, 
if EOS in the outer crust is known.
But how can we determine $P_\mathrm{oi}$?
In Ref.\ \cite{gc20} 
we found $P_\mathrm{oi}$ for the smoothed CLD model based on the SLy4 
nuclear energy-density functional \cite{Chabanat_ea98_SLY4}.
Here, to check the sensitivity of $P_\mathrm{oi}$ 
to the shell effects, 
we determined  
it
for the recently developed model 
\cite{carreau_ea20_cryst_inner_CLDM}. 
The shell effects in this model are incorporated 
by adding tabulated 
shell energy corrections
from \cite{Pearson_ea18,Pearson_ea19_erratum_tables} on top of 
CLD energy density (CLD+sh model in what follows). The resulting CLD+sh model reproduces well 
ETFSI calculations for four modern energy-density functionals: BSk22, BSk24, BSk25, and BSk26.
As in the most advanced  calculations to date 
performed in the
traditional approach \cite{Fantina_ea18},
we assume that the inner crust consists of nuclei 
of one particular species at each given density.
A detailed discussion of our results 
based on the CLD+sh model \cite{carreau_ea20_cryst_inner_CLDM}
is presented in \cite{gc20_large}.
Here we describe our basic findings, 
obtained for pure $A=56$ composition of the ashes and one of the 
CLD+sh models of \cite{carreau_ea20_cryst_inner_CLDM}, corresponding  to BSk24 functional.
The results for other functionals are similar.

(i) As in the CLD model of \cite{gc20},
for CLD+sh model \cite{carreau_ea20_cryst_inner_CLDM} 
there exists 
a pressure $P_\mathrm{oi}^{\rm (min)}$ such that, for 
any $P_\mathrm{oi} \geq P_\mathrm{oi}^{\rm (min)}$ 
the construction of the AC model is limited 
by the instability 
disintegrating nuclei into neutrons.
The pressure $P_{\rm inst}(P_\mathrm{oi})$, 
at which this instability takes place, 
is a decreasing function of $P_\mathrm{oi}$.
Thus, 
$P_{\rm inst}(P_\mathrm{oi}^{\rm (min)})$
is a maximum possible value of $P_{\rm inst}$.
Generally, disintegration of nuclei
is accompanied by the energy release.
It is interesting to note that for the CLD+sh model \cite{carreau_ea20_cryst_inner_CLDM}, 
the disintegration instability (for any $P_{\rm oi}$) occurs at 
$P_{\rm inst}$
smaller than the pressure at the crust-core boundary 
(where both $P$ and $\mu_n$ must be matched). 
A part of the crust at $P>P_{\rm inst}$ 
appears to be decoupled from the rest of the crust:
the atomic nuclei in this ``relic'' part
are not replaced during the accretion in the FAC regime 
(because all the upcoming nuclei disintegrate at $P=P_{\rm inst}$).
The relic part of the crust is formed
during the transformation from the pristine catalyzed crust to FAC.

(ii) The shell effects
complicate
determination of $P_\mathrm{oi}$ in FAC. 
This happens because the relic part of the crust
can be (at least, in principle) 
stabilized by the shell effects for a range of $P_{\rm oi}$.
The reason for that is largely unknown composition of nuclei in the relic region,
which depends on 
the 
(highly uncertain) 
evolution preceding the FAC formation.
Thus, unambiguous determination of $P_\mathrm{oi}$
remains a task for the future.
However, 
we numerically found that for
$A=56$ ashes and CLD+sh model \cite{carreau_ea20_cryst_inner_CLDM}, $P_\mathrm{oi}$
does not exceed $P_{\rm nd}^\mathrm{(cat)}$: 
for higher $P_\mathrm{oi}$ the pressure $P_\mathrm{inst}$ 
becomes so low, 
that FAC at $P<P_\mathrm{inst}$ and NS core can not be connected in a thermodynamically consistent way for any composition of the relic part of the crust \cite{gc20_large}.
The latter result 
is obtained 
assuming 
that shell corrections can be ignored for baryon densities larger than 
the proton drip density, which is chosen to be
$0.073$~fm$^{-3}$, the same as in \cite{Pearson_ea18}.

Obviously, 
the lower bound on $P_\mathrm{oi}$ 
equals
the minimal value of $P_\mathrm{oi}$, at which the instability takes place, 
i.e., $P_\mathrm{oi}=P_\mathrm{oi}^{({\rm min})}$.
According to our calculations, for the CLD+sh model considered here, we have 
$P_\mathrm{oi}^{\rm (min)}\approx 0.91 P_{\rm nd}^{\rm (cat)}$,
$P_{\rm inst}(P_{\rm oi}^{\rm (min)})\approx 0.267$~MeV fm$^{-3}$, and 
$Q_i(P_{\rm oi}^{\rm (min)}) \approx 0.06$~MeV/nucleon.
Independently,  $P_\mathrm{oi}$ can be bounded from below by the condition 
$Q_\mathrm{i}>0$ (see Fig.\ \ref{Fig_Q}): 
otherwise
disintegration of nuclei 
is not energetically favorable and can not proceed. 
The $P_{\rm oi}$ value corresponding to $Q_{\rm i}=0$
is denoted as 
$P_\mathrm{oi}^{(0)}$.
Clearly, it must be 
$P_\mathrm{oi}^{(0)}\leq P_\mathrm{oi}^{\rm (min)}$. 	 
For the CLD+sh model
$P_\mathrm{oi}^{(0)}\approx 0.89 P_{\rm nd}^{\rm (cat)}$ 
and is close to $P_\mathrm{oi}^{({\rm min})}$, 
so that below
we consider $P_\mathrm{oi}^{(0)}$ as a universal lower bound on $P_{\rm oi}$.

(iii) Assuming pure $A=56$ composition of the ashes, 
the charge number of nuclei at the bottom of the outer crust is $Z=20$.
The shell effects stabilize $Z$ at 
the value $Z=20$
in 
almost whole region $P\leq P_{\rm inst}(P_\mathrm{oi})$.
This result is 
related to the 
local energy minimum at $Z=20$ 
(which is the proton `magic number' 
in the inner crust \cite{Pearson_ea18}) 
and does not depend on the choice of $P_\mathrm{oi}$. 
Constancy of $Z$ in the inner crust implies that almost all heat $Q_\mathrm{i}$ 
is released at the instability point
$P=P_{\rm inst}(P_\mathrm{oi})$.
(Let us note that the heat release in the inner crust can not be associated 
with the change of the mass number $A$, which is 
treated as a continuous variable 
due to the presence of unbound neutrons 
\cite{Fantina_ea18,Pearson_ea19_erratum_tables, carreau_ea20_cryst_inner_CLDM}.)

\begin{table}
	\begin{tabular}{llccccc}
		\hline
		\hline
		Model& $P_\mathrm{oi}$ &   $P_\mathrm{oi}/P_\mathrm{nd}^\mathrm{(cat)}$  &$Q_\mathrm{o}$ &$Q_\mathrm{oi}$&$Q_\mathrm{i}$&$Q$\\
				\hline
		FRDM12\ \ & $P_\mathrm{oi}^\mathrm{(0)}$
		& 0.85
		&0.18\ \ & 0.03\ \ & 0.00\ \ & 0.21\\
		&$P_\mathrm{nd}^\mathrm{(cat)}$
		&1.00
		&0.18& 0.03& 0.32 & 0.53\\
		\hline
		BSk24& $P_\mathrm{oi}^\mathrm{(0)}$
		& 0.92 
		&0.12& 0.17& 0.00 & 0.29\\
		&$P_\mathrm{nd}^\mathrm{(cat)}$
		&1.00
		&0.12& 0.15& 0.19 & 0.46\\
		\hline
		BSk25& $P_\mathrm{oi}^\mathrm{(0)}$
		& 0.93 
		&0.17& 0.19& 0.00 & 0.36\\
		&$P_\mathrm{nd}^\mathrm{(cat)}$
		&1.00
		&0.17& 0.17& 0.17 & 0.51\\
				\hline
		BSk26& $P_\mathrm{oi}^\mathrm{(0)}$
		& 0.96 
		&0.14& 0.25& 0.00 & 0.39\\
		&$P_\mathrm{nd}^\mathrm{(cat)}$
		&1.00
		&0.14& 0.13& 0.20 & 0.47\\
		\hline \hline
	\end{tabular}
	\caption{Heat release distribution for the limiting values of $P_\mathrm{oi}$.
	$Q$-values are in MeV/nucleon.}
	\label{Tab:heat}
\end{table}

Table \ref{Tab:heat} represents the heat release distribution 
in FAC for two values of $P_\mathrm{oi}$: 
the lower bound $P_\mathrm{oi}^{(0)}$($\approx P_{\rm oi}^{({\rm min})}$),
at which $Q_\mathrm{i}=0$ and for $P_\mathrm{oi}=P_\mathrm{nd}^\mathrm{(cat)}$, 
which 
bounds
$P_\mathrm{oi}$ from above.
As in the case of Fig.\ \ref{Fig_Q},
Tab.\ \ref{Tab:heat} was calculated using 
the mass tables \cite{FRDM12,HFB24},
rather than the simplified CLD+sh model based on ETFSI calculations.
These two approaches lead to a bit different predictions 
(in particular, $P_{\rm nd}^{\rm (cat)}$ differs by 
a few percent \cite{Pearson_ea18}),
which explains 
why 
$P_{\rm oi}^{(0)}=0.92 P_{\rm nd}^{\rm (cat)}$ 
for BSk24 model in Tab.\ \ref{Tab:heat}
appears to be larger than $P_{\rm oi}^{\rm (min)}=0.91 P_{\rm nd}^{\rm (cat)}$,
calculated employing CLD+sh model.

As follows from Fig.\ \ref{Fig_Q}, 
the minimal heat release 
$Q$ occurs if $P_\mathrm{oi}$ equals the minimal value, 
at which disintegration instability takes place, $P_\mathrm{oi}=P_\mathrm{oi}^{\rm (min)}$.
It is interesting to point out, that exactly this $P_\mathrm{oi}$ should be realized in FAC, 
if the Prigogine Minimum Entropy Production theorem \cite{prigogine_IntrodNEqThermod}
works for
our problem.

\textit{Summary and conclusions.--}
We present a universal formula (\ref{Q_dch}) 
for the heat release
$Q^\infty$
in the 
fully accreted NS 
crust by nonequilibrium nuclear reactions.
The formula is applicable to {\it arbitrary} composition of nuclear ashes
and crust model.
We further
analyze the heat release 
in the outer crust $Q_{\rm o}$, in the inner crust $Q_{\rm i}$, 
and at the oi interface, $Q_\mathrm{oi}$, 
for thermodynamically consistent FAC model
respecting the nHD condition. 
We show that these quantities 
are parametrized
by the pressure $P_\mathrm{oi}$ at the oi interface,
provided that the nuclear mass model in the outer crust is specified 
[see Fig.\ \ref{Fig_Q} and Eq.\ (\ref{Q_dch2_nHD})].
To calculate  $P_\mathrm{oi}$ for FAC and determine the heat release distribution $Q_{\rm i}$ in the inner crust, 
we apply CLD+sh model of \cite{carreau_ea20_cryst_inner_CLDM}
with shell corrections. 
We demonstrate that 
for 
the ashes composed of $^{56}$Fe,
\textit{almost all heat $Q_{\rm i}$ 
is released at the instability point}, 
where nuclei disintegrate into neutrons  (at $P=P_{\rm inst}$).
This result does {\it not} depend on the actual value of $P_\mathrm{oi}$.
The charge number is {\it fixed} at the value $Z=20$
in almost whole region 
between the oi interface and instability point.

We also argue that account for shell effects
complicates unambiguous determination of $P_\mathrm{oi}$,
which then depends on the way the FAC is formed.
Our analysis indicates that $P_{\rm oi}$ for the employed CLD+sh model 
\cite{carreau_ea20_cryst_inner_CLDM} 
and $^{56}$Fe ashes 
is bounded from below by
$P_\mathrm{oi}^{\rm (min)}$
-- the minimal possible value of $P_\mathrm{oi}$, 
for which the disintegration instability can occur in the inner crust
\cite{gc20}.
In turn, we also numerically found 
that 
$P_{\rm oi}$ 
does not exceed
$P_{\rm nd}^{({\rm cat})}$ -- the neutron drip pressure
in the catalyzed crust (see \cite{gc20_large} for more details).
We emphasize that this upper bound on $P_{\rm oi}$
is model-dependent, being sensitive to the 
density profiles
of the shell energies,
which are still 
poorly known
at large baryon densities \cite{Pearson_ea18,pcp20}.

As a conservative estimate of $P_\mathrm{oi}$ in the FAC regime, 
we propose to take
$P_\mathrm{oi}^{\rm (0)}\leq P_\mathrm{oi}<P_{\rm nd}^\mathrm{(cat)}$, 
where $P_\mathrm{oi}^{(0)}$ corresponds to $Q_\mathrm{i}=0$,
and is just a bit smaller than $P_\mathrm{oi}^{\rm (min)}$
according to our calculations.
The respective heat release distribution is shown in Tab.\ \ref{Tab:heat} (see also Fig.\ \ref{Fig_Q}).
In particular, 
the deep crustal heating energy release 
$Q^{\infty} \sim (0.21- 0.53) e^{\nu_\mathrm{oi}/2}$~MeV/nucleon,
appears to be at least several times smaller than in the traditional approach, 
$Q^{\infty} \sim (1.5-2) e^{\nu_\mathrm{oi}/2}$~MeV/nucleon.
This fact
calls for the reinterpretation of the existing observational data
on thermal properties of transiently accreting NSs
and should stimulate further work on developing realistic accreted crust models.
 
Concluding, we
stress that
the information 
provided in this Letter is, in principle, sufficient to start modeling 
the thermal relaxation 
of X-ray transients,
as well as their quiescent temperatures 
within the nHD approach.
Such a modeling may help further constrain the pressure $P_\mathrm{oi}$.

\begin{acknowledgments}
\textit{Acknowledgments.--}
MEG is grateful to E.M.~Kantor for useful discussions 
and acknowledges support by 
the Foundation for the Advancement of Theoretical Physics and mathematics 
``BASIS'' [Grant No.\ 17-12-204-1] and by RFBR [Grant No.\ 19-52-12013]. 
\end{acknowledgments}


%

\newpage

\begin{cbunit}
	
	\renewcommand{\thesection}{\arabic{section}}
	\renewcommand{\thesubsection}{\arabic{section}.\arabic{subsection}}
	\renewcommand{\section}[1]
	{
		\refstepcounter{section}
		\begin{center}
			{ \bf \thesection.\ #1}
		\end{center}
	}
	\renewcommand{\subsection}[1]
	{
		\refstepcounter{subsection}
		\begin{center}
			{\it \thesubsection.\ #1}
		\end{center}
	}

	\onecolumngrid
	\begin{center}
		{\bf \Large Supplemental material}
	\end{center}

In section \ref{Sec_mub} of the Supplemental material 
we derive an estimate for $\partial M/\partial A_b$,
which justifies the 
second line of equation (3) in the Letter.
Section \ref{Sec:tradmod} demonstrates that the general equation (3) 
from the Letter 
reproduces  
(and refines) 
previous calculations of the heat release performed in the traditional approach. 
In section \ref{Sec:alt_heat} we derive a formula for the deep crustal heat release, 
which is equivalent to equation (4) of the Letter.
This formula is suitable for estimating 
sum of the heat releases at the outer-inner  crust interface and in the inner crust. 
Finally, section \ref{Sec:heat_CLDM} contains a microscopic derivation 
of the heat release for the smoothed compressible liquid drop (CLD) model within 
the thermodynamically consistent approach.
It is shown that the result is in agreement with our general formula (3)
from the Letter.

\section{$\partial M/\partial A_b$ for neutron stars with fully accreted crust} \label{Sec_mub}

We consider a family of neutron stars (NSs) 
with fully accreted crust (FAC), 
parametrized by the total baryon number in the star, $A_b$.
We neglect temperature effects and assume that all configurations 
in the family have the same equation of state (EOS),
i.e., the pressure $P$, energy density $\epsilon$, 
baryon chemical potential $\mu_b\equiv (P+\epsilon)/n_b$,
and composition 
are some definite functions  
of the baryon number density, $n_b$.  
We also assume that the NS core is in the ground state,
while FAC is not.

The gravitational mass $M$ and $A_b$ of a FAC NS
are given by \cite{hpy07}:
\begin{eqnarray}
	M&=&4\pi \int \epsilon r^2 \mathrm d r, \label{MFA} \\
	A_b&=&4\pi \int n_b e^{\lambda/2} r^2 \mathrm d r, \label{Ab}
\end{eqnarray}
where $r$ and $\lambda$ are the radial coordinate and 
spatial metric coefficient, 
respectively.

For NSs with the ground-state (catalyzed) composition in the crust and core,  
the derivative $d M_\mathrm{cat}/d A_b$ is well known (e.g., \cite{ZN71,hpy07}), 
\begin{equation}
	\frac{d M_\mathrm{cat}}{d A_{b}}=\mu_{b}^\infty,
	\label{mueq}
\end{equation}
where $M_\mathrm{cat}$ is the mass of a catalyzed NS; 
$\mu_{b}^\infty=\mu_b e^{\nu/2}$ is the redshifted baryon chemical potential, 
which is constant in the catalyzed star; 
and $\nu$ is the temporal metric coefficient.
However, since FAC 
is not in the ground state,
Eq.\ (\ref{mueq}) does not hold,
in particular,  
$\mu_{b}^\infty$ varies in the crust \cite{zfh17,gc20}.

To determine $\partial M/\partial A_b$ for an NS with FAC
let us note, that  the functions $\epsilon(r)$, $\nu(r)$, $\lambda(r)$ 
can be found from the Tolman-Oppenheimer-Volkoff (TOV) equations 
(see, e.g., \cite{hpy07}), 
which depend {\it only} on the thermodynamic function $P(\epsilon)$.
This property allows us to use a simple trick that considerably simplifies subsequent calculations.
Namely,
consider some 
fictitious particles
with the same $P(\epsilon)$ as in NSs with FAC
and
assume that it is
the {\it ground state} EOS for them.
Introducing the
chemical potential $\mu_f$ and number density $n_f$ 
for these particles, one can write
\begin{align}
	&
	d \epsilon =\mu_f dn_f,
	&
	\label{1}\\
	&
	P+\epsilon=\mu_f n_f.
	&
	\label{2}
\end{align}
Using these equations together with the one of the TOV equations, 
$dP/dr=-[(P+\epsilon)/2] \,  d\nu/dr$,
it is straightforward to verify that $\mu_f^\infty=\mu_f e^{\nu/2}={\rm const}$
in the star, i.e., it is indeed the ground state for the fictitious particles.
Because in the NS core (which is in the ground state) 
the same equations (\ref{1}) and (\ref{2}) are valid with $\mu_f$
replaced by
$\mu_b$ and
$n_f$ replaced by $n_b$, 
one can always choose $\mu_f$ and $n_f$ in such a way 
that at any point in 
the core: $\mu_{f}=\mu_{b}$, $n_f=n_b$.
Summarizing, 
\begin{align}
	\underline{\rm Everywhere}: \,\,\, \mu_f^\infty={\rm const}, 
	\quad \quad \underline{\rm 
		At\ any\ point\ in\ the\ core}: \,\,
	\mu_{f}=\mu_{b}, \quad n_f=n_b.
	\label{muf}
\end{align}

Because fictitious particles are in the ground state, Eq.\ (\ref{mueq})
is satisfied
for them as well
\begin{equation}
	\frac{d M}{d A_f}=\mu^\infty_f=\mu^\infty_{b,\mathrm{core}},
	\label{dMfdAf}
\end{equation}
where
\begin{equation}
	A_f=4\pi \int n_f e^{\lambda/2} r^2  d r
	\label{Af}
\end{equation}
is the total number of fictitious particles in the star
and in Eq.\ (\ref{dMfdAf}) we denoted the (constant) redshifted baryon chemical potential in the core
as 
$\mu_{b,{\rm core}}^\infty$
(the same notation is used in the Letter).

Clearly, in the crust of a FAC NS $\mu_b \neq \mu_f$.
Assuming that the difference $\Delta \mu_b(P) \equiv \mu_b(P)-\mu_f(P) \ll \mu_f(P)$
is small (the subsequent derivation will confirm this), one 
can expand the baryon number density as
[recall that $\mu_b=(P+\epsilon)/n_b$]
\begin{equation}
	n_b(P)=\frac{P+\epsilon}{\mu_b}\approx n_f(P)\,
	\left[1-\frac{\Delta \mu_b(P)}{\mu_f(P)}\right].
\end{equation}
Then, accounting for (\ref{Ab}), we can write
\begin{eqnarray}
	A_b
	&\approx& A_f- 4\pi \int n_f 
	\frac{\Delta \mu_b(P)}{\mu_f(P)} e^{\lambda/2} r^2 \mathrm d r
	=  A_f-\mathcal{O} \left(\frac{\Delta \mu_{\rm typ}^\infty}{\mu_f^\infty} A_c\right).
	\label{A_b_approx}
\end{eqnarray}
As far as $\Delta \mu_b(P)=0$ in the core [see Eq.\ (\ref{muf})], 
only crust contributes to the integral here,
thus justifying the last equality. 
In Eq.\ (\ref{A_b_approx}) 
$\mathcal{O}(\ldots)$ denotes an order of magnitude;
$A_c$ is the number of baryons in the crust;
and $\Delta \mu_{\rm typ}^\infty$ represents the typical value of $\Delta \mu_b^\infty$ in the crust.
$\Delta \mu_{\rm typ}^\infty$ can be estimated as:
\begin{align}
	&
	\Delta \mu_{\rm typ}^\infty \sim \left. \mu_b^\infty\right|_{r=R} -\mu_{f}^\infty=
	\overline{m}_b e^{\nu_s/2}-\mu_{b,{\rm core}}^\infty, 
	&
	\label{dmutyp}
\end{align}
where
$\left. \mu_b^\infty\right|_{r=R}=\overline{m}_b\,e^{\nu_s/2} $
is the redshifted
baryon chemical potential at the stellar surface ($r=R$);
if measured locally, it equals the average mass per baryon, $\overline{m}_b$
(see the Letter);
$\nu_s$ is the redshift factor at the surface; and
we used 
Eq.\ (\ref{dMfdAf}): 
$\mu_f^\infty=\mu_{b,{\rm core}}^\infty$.

Now, we can apply (\ref{A_b_approx}) to calculate $\partial M/\partial A_b$
at fixed FAC EOS:
\begin{equation}
	\frac{\partial M}{\partial A_b}
	=
	\frac{d M}{d A_f}
	\frac{d A_f}{d A_b}
	= \mu_{b,\, \mathrm{core}}^\infty\,\left[1+\frac{\partial}{\partial A_b}\, 
	\mathcal{O} \left(\frac{\Delta \mu_{\rm typ}^\infty}{\mu_f^\infty} A_c\right) \right]
	\approx
	\mu_{b,\, \mathrm{core}}^\infty+
	\Delta \mu_{\rm typ}^\infty\,
	\mathcal{O}\left( \frac{
		A_c}{
		A_b}
	\right),
	\label{FinEstim}
\end{equation}
where the last 
approximate equality is obtained assuming no 
strong phase transitions in the stellar centre 
(at least, in the vicinity of $A_b$). 
Consequently, $A_c$ is a smooth function of $A_b$,
so that
\begin{equation}
	\frac{d 
		A_c}{d
		A_b}
	=\mathcal{O}\left( \frac{
		A_c}{
		A_b}
	\right)
	=\mathcal{O}\left( \frac{
		M_c}{
		M}
	\right).
\end{equation}
As long as the mass of the crust 
$M_c\approx m_{U} A_c\sim 10^{-2} M$ 
($m_{U}$ is the atomic mass unit), the last term in the right-hand side of Eq.\ (\ref{FinEstim}) 
is much smaller than the first one.
Thus, recalling equation (2) from the Letter and using Eq.\ (\ref{dmutyp}), Eq.\ (\ref{FinEstim})
can be rewritten in the required form
\begin{equation}
	\frac{\partial M}{\partial A_b}
	=
	\mu_{b,\, \mathrm{core}}^\infty+
	\mathcal{O}\left( \frac{
		Q^\infty_{\rm tot} M_c}{
		M}
	\right).
	\label{FinEstim2}
\end{equation}
Here $Q^\infty_{\rm tot}$ is the total heat release per accreted baryon, 
redshifted to a distant observer.

\section{Deep crustal heat release in the traditional approach}
\label{Sec:tradmod}

In the traditional approach 
one follows the evolution of the compressed matter element under the increasing pressure $P$,
allowing for electron captures, neutron emissions and pycnonuclear
reactions,
but neglecting possible redistribution of unbound neutrons
within the 
inner crust.
In the literature, it is often assumed that the nuclear burning ashes are 
composed of $^{56}$Fe and, furthermore, the accreted crust 
has
an ``onion'' structure, 
i.e.\ 
consists of a number of layers 
in which nuclei of only one particular species are present
(one-component approach, \cite{HZ90,HZ90b,HZ03,HZ08,Fantina_ea18}).
To compare our general approach 
[leading to equation (3) of the Letter] with these works,
we adopt the same simplifications in this section.

For $^{56}$Fe ashes the deep crustal heat release $Q^\infty$, 
given by the equation (3) of the Letter, can be rewritten as
\begin{align}
	&Q^\infty = (m_{\rm Fe}/56)\,e^{\nu_{s}/2}-
	\mu_{b, {\rm core}}^\infty,&
	\label{Q_dch2}
\end{align}
where we made use of the fact that the average mass per baryon in the ashes equals
$\overline{m}_{b,{\rm ash}}=m_{\rm Fe}/56$ ($m_{\rm Fe}$ is the mass of $^{56}$Fe) 
and also neglected small corrections 
$\sim Q^\infty_{\rm tot} M_c/M \ll 1$~MeV.

Within 
the one-component approach,
the neighbouring layers are separated by interfaces $i$ with the pressure $P_i$,
corresponding to   
thresholds of nuclear reactions, 
where the composition changes  abruptly. 
At $P=P_i$
the baryon chemical potential
exhibits jumps,
which leads to the
heat release 
$Q_i=\mu_i^- -\mu_i^+$ 
per each
baryon, crossing the $i$-th interface \cite{HZ08,zfh17}
($\mu_i^-$ and $\mu_i^+$ are, respectively, 
$\mu_b$ 
immediately before and after the jump).
Ref.\ \cite{zfh17} discussed the thickness of FAC within 
the traditional one-component approach (assuming $^{56}$Fe ashes),
and suggested a set of formulas similar (but not equivalent) to (\ref{Q_dch2}).
Here we demonstrate how to modify the consideration in \cite{zfh17} 
in order to reproduce our formula (\ref{Q_dch2}).

The deep crustal heat release $Q$ in \cite{zfh17} was calculated as $Q= \sum_i Q_i$, 
i.e., as a sum over the local heat release sources, 
without accounting for the redshift factors, $e^{\nu_i/2}$. 
However, these factors 
are not exactly the same for all heat sources, 
simply because they are 
taken at 
different
depths
in the crust.
This effect can be easily 
handled
if we calculate 
$Q^\infty$ -- the deep crustal heat release
redshifted to a distant observer:
$Q^\infty=\sum Q^\infty_i=\sum_i Q_i\,e^{\nu_i/2}$. 
As long as
$\mu_b^\infty$ is constant in the layers between the interfaces 
(for $P_{i}<P<P_{i+1}$) \cite{zfh17},  
$\mu_{i+1}^-=e^{(\nu_{i}-\nu_{i+1})/2}\mu_{i}^+$. 
Thus, all intermediate terms in 
the sum cancel out. 
Eventually, we arrive at the formula: 
$Q^\infty=\mu_0^\infty-\mu_L^\infty$, 
where $\mu_0^\infty$ and $\mu_L^\infty$ 
are, respectively, 
the redshifted
$\mu_b$
before the first and after the last  interface, 
i.e.\ for the ashes at the surface and for the matter in the core. 
As a result, $Q^\infty$ coincides with our Eq.\ (\ref{Q_dch2}).

In more realistic versions of the traditional approach, 
the one-component approximation should be relaxed.
Then the heat is released not at the interfaces $P=P_i$, 
but in some pressure regions 
(e.g., \cite{lau_ea18,SC19_MNRAS,SC19_JPCS}). 
To calculate $Q^\infty$ in this case,
one should integrate the local heat release, 
accounting for the variation of the redshift factor throughout the crust. 
However, it is easier to determine $Q^\infty$ directly from Eq.\
(3) of the Letter, 
which can be used if FAC EOS is known.

\section{An alternative formula for the heat release in the thermodynamically consistent model } 
\label{Sec:alt_heat}

Equation (4) of the Letter 
represents the general and accurate formula for the deep crustal heat release 
calculated
for FAC within the thermodynamically consistent approach, 
which respects the neutron hydrostatic/diffusion equilibrium
(nHD) condition.
The latter condition states that the redshifted
neutron chemical potential $\mu_n^\infty$ must be constant in the inner crust:
$\mu_n^\infty=\mu_n e^{\nu/2}=\mathrm{const}$.
Here we derive 
an alternative form of that equation, 
which explicitly separates the heat release in the outer crust, 
$Q_{\rm o}^\infty$, and  the remaining sum of
the heat releases in the inner crust, $Q^\infty_\mathrm i$, and
at the outer-inner crust interface (oi interface), $Q^\infty_\mathrm{oi}$.
Neglecting small corrections 
$\sim Q^\infty_{\rm tot} M_c/M \ll 1$~MeV 
and making use of the fact that $\mu_{\rm b, {\rm core}}^\infty=m_n e^{\nu_\mathrm{oi}/2}$
(see the Letter for details), 
where $m_n$ is the neutron mass and
$\nu_\mathrm{oi}$ is the redshift factor at the oi interface,
equation (3) from the Letter can be rewritten as
\begin{eqnarray}
	Q^\infty&=&
	\overline{m}_{b,{\rm ash}}\,e^{\nu_{ s}/2}-
	m_n e^{\nu_\mathrm{oi}/2} 
	=\underline{\overline{m}_{b,{\rm ash}}\,e^{\nu_{ s}/2}
		-\mu_{b,\mathrm{oi}}^- e^{\nu_\mathrm{oi}/2}}
	+(\mu_{b, \mathrm{oi}}^- -m_n) e^{\nu_\mathrm{oi}/2}.
	\label{import}
\end{eqnarray}
Here
$\mu_{b, \mathrm{oi}}^-$ is the baryon chemical potential at the bottom 
of the outer crust, 
which can be calculated,
following the traditional approach,
as a function 
of the pressure $P_{\rm oi}$ at the oi interface.
Note that $e^{\nu_{s}/2}$ cannot be replaced with $e^{\nu_\mathrm{oi}/2}$ 
in Eq.\ (\ref{import}): this would significantly affect $Q^\infty$.
The underlined terms in Eq.\ (\ref{import}) represent the (redshifted) 
heat release in the outer crust, $Q_{\rm o}^\infty$ 
(c.f.\ the derivation 
of $Q^\infty$ in the traditional one-component approach in section \ref{Sec:tradmod}).
The remaining term determines 
the sum $Q^\infty_\mathrm{oi}+Q^\infty_\mathrm{i}$.

\section{Heat release for nHD crust within the smoothed CLD model} 
\label{Sec:heat_CLDM}
As shown in \cite{gc20}, 
the thermodynamically consistent model of the inner crust should respect the nHD condition.
To account for nHD condition, one should 
substantially modify the traditional approach:
Instead of studying nuclear transformations 
occurring (due to compression)
in a particular element of accreted matter, 
one should self-consistently analyze nuclear processes in the {\it whole} inner crust, 
allowing for
redistribution of unbound neutrons over different crust layers and the core. 
This problem was addressed in \cite{gc20} within the smoothed CLD model, 
in which nuclei are described as liquid drops, 
located at the centre of the spherical Wigner-Seitz  cells \cite{lpr85,hpy07,ch08}.  
Here we calculate the deep crustal heat release $Q^\infty$ 
for this model within the microscopic approach and demonstrate 
that it agrees with the general result, 
presented in the Letter  [see equation (3) there].

\subsection{Basic features of FAC within the CLD model}

Within the smoothed CLD model charge and atomic mass numbers, $Z$ and $A$, 
are treated as continuous variables;
pairing and shell effects are ignored.
We also assume that 
mechanical, diffusion, and beta-equilibrium
conditions
hold true 
inside each Wigner-Seitz  cell (see Supplemental material in \cite{gc20} for details)
and neglect finite-temperature effects.
Using these conditions  the energy density 
can be presented 
as a two-parameter function, 
$\epsilon(n_b, n_N)$,
where $n_N$ is the number density of nuclei,
which is not the sole function of $n_b$ (as in the catalyzed crust),
because  
temperature in the crust is too low to initiate
nuclear reactions required 
to drive $n_N$ to its equilibrium value. 
The second law of thermodynamics can be written as \cite{gc20}:
$d \epsilon(n_b, n_N)=\mu_n d n_b+\mu_N d n_N$,  
where the effective chemical potential $\mu_{N}$ describes the energy
change due to 
creation of additional
nuclear cluster in the
system at fixed $n_b$. 

The outer crust can be considered within the traditional approach with the constant $A$, 
because unbound neutrons there are absent and none of the nuclear reactions can change $A$. 
Thus, inside the outer crust the number density of nuclei is
$n_N=n_b/A$, so that 
$d\epsilon= (\mu_n+\mu_N/A) d n_b=\mu_b d n_b$, 
where $\mu_b=(P+\epsilon)/n_b=(\mu_n+\mu_N/A)$. 
Then the Gibbs-Duhem relation reads:
$dP=n_b\,d\mu_b$.
Combining this relation with the hydrostatic equilibrium condition (one of the TOV equations), 
\begin{equation}
	\frac{dP}{dr}=-\frac{1}{2}(P+\epsilon) \frac{d\nu}{dr},
	\label{hydrostat}
\end{equation}
we find
\begin{equation}
	\frac{d \mu_b}{dr}=-\frac{\mu_b}{2}\frac{d\nu}{dr},
	\label{mu_b_const}
\end{equation}
i.e., $\mu_{b}^\infty={\rm const}$. 

For the inner crust,
a combination of 
the nHD condition, $\mu_n^\infty=\mu_n e^{\nu/2}=\mathrm{const}$,
with the hydrostatic equilibrium equation (\ref{hydrostat}), 
and 
the Gibbs-Duhem relation, 
$dP= n_b\, d\mu_n+n_N\, d\mu_N$,
leads to an additional condition, $\mu_N^\infty=\mu_N e^{\nu/2}=\mathrm{const}$, 
which must be fulfilled in the nHD crust \cite{gc20}.
Correspondingly, the ratio $C \equiv \mu_N/\mu_n$
is constant in the inner crust
and 
parametrizes the {\it family of nHD EOSs}.

In the important case of FAC the parameter $C$ 
(and hence FAC EOS) is fixed
by the requirement that 
the disintegration 
instability exists, 
which allows to convert nuclei into 
neutrons. For SLy4 model considered in \cite{gc20} it 
occurs at the bottom of the crust.
As it should be,
$P$ and $\mu_n$ are continuous across the crust-core boundary (cc boundary).
In turn, at the oi interface the pressure $P_{\rm oi}$ is continuous, 
while the chemical potential of unbound neutrons at the top of the inner crust equals 
$\mu_{n,  {\rm oi}}^+=m_n$ ($m_n$ is the bare neutron mass).
Here and below, for any quantity $X$ the notations $X_{\rm oi}^-$ and $X_{\rm oi}^+$
mean that $X$ is taken at the oi interface from the outer 
and the inner 
crust sides, respectively.

\subsection{Heat release within the smoothed CLD model: microscopic derivation}
Within the smoothed CLD model
$Z$ and $A$ vary smoothly in the crust 
(see Fig.\ 2 in \cite{gc20}) 
and there is no heat release 
except for
the phase transitions 
at: (i) the oi interface
and
(ii) the cc boundary,
where nuclear transformations proceed in a strongly nonequilibrium regime.
Let us first calculate heat release at these 
phase transitions
microscopically.
To do that, it 
is useful to introduce 
the concept of chemical potential of a 
nucleus, $\mu_{\mathrm{nuc}}$,
which is equal to the change of the system energy 
if a nucleus 
is added to the system.
For the CLD model: 
$\mu_{\mathrm{nuc}}=A\mu_n+\mu_N=A\mu_\mathrm b$
(see expression for $d\epsilon$ above).

Let us start with the oi interface. 
At the interface, a 
nucleus
from the outer crust with $A=A_{\rm oi}^-$
enters
the inner crust, absorbs $\Delta A=A_{\rm oi}^+-A_{\rm oi}^-$ 
unbound neutrons with $\mu_n=\mu_{n,{\rm oi}}^+=m_n$, 
and becomes 
a nucleus
in the inner crust with $A=A_{\rm oi}^+$. 
The heat release $\mathcal{Q}_\mathrm{oi}$ for this process 
(per nucleus) 
is given by the difference 
between the sum of chemical potentials before and after the reaction:
$\mathcal{Q}_\mathrm{oi}=\mu^-_{\mathrm{nuc,oi}} + \Delta A \mu_{n,{\rm oi}}^+-\mu_\mathrm{nuc,oi}^+
=A_{\rm oi}^- \mu_{b,{\rm oi}}^- -A_{\rm oi}^- \mu_{n,{\rm oi}}^+ -\mu_{N,{\rm oi}}^+$.
In turn, at the cc boundary 
a nucleus
(with the nucleon number $A_\mathrm {cc}$) 
disintegrate into 
neutrons because of the instability discussed in \cite{gc20},
leading to the heat release
$\mathcal{Q}_\mathrm {cc}=\mu_\mathrm{nuc,cc}-A_\mathrm {cc} \mu_{n,\mathrm{cc}}
=\mu_{N,{\rm cc}}$, where 
the subscript ${\rm cc}$ denotes the corresponding quantity at the boundary.

In the regime when accretion occurs onto an NS with the FAC,
the total amount of nuclei in the inner crust remains almost constant 
[small deviations, 
associated with the change of 
radius, gravity and metric functions 
in the course of accretion,
lead to the effects $\sim Q_{\rm tot}^\infty M_{c}/M$ 
on the heat release $Q^\infty$ and can be ignored, 
see the similar discussion in section \ref{Sec_mub}].
Thus,
the number of nuclei (per unit time), 
crossing the oi interface should coincide with the number of nuclei crossing cc boundary.
Correspondingly, 
\begin{equation}
	Q^\infty=
	\frac{1}{A_{\rm{oi}}^-} 
	\left(\mathcal{Q}_\mathrm{oi} e^{\nu_\mathrm{oi}/2}
	+\mathcal{Q}_\mathrm{cc} e^{\nu_\mathrm{cc}/2}\right)
	=\overline{m}_{b,\mathrm{ash}} e^{\nu_s/2}-\mu_{b, {\rm core}}^\infty,
	\label{Q4}
\end{equation}
being in agreement with the general Eq.\ (3) in the Letter. 
To obtain the last equality we noted that:
(i) $\mu_{N}^\infty$ and $\mu_n^\infty$ 
are both constants in the inner crust 
and $\mu_n^\infty= \mu_{b, {\rm core}}^\infty$;
(ii) For CLD model $\mu_{b}^\infty$ is constant in the outer crust [Eq.\ (\ref{mu_b_const})], hence
$\mu_{b, {\rm oi}}^- e^{\nu_{\rm oi}/2}=\overline{m}_{b,{\rm ash}}e^{\nu_{ s}/2}$.
In Eq.\ (\ref{Q4}) 
$\nu_\mathrm{cc}$ is the redshift
at the 
cc boundary.

The above discussion is applicable to arbitrary smoothed CLD model. 
Let us provide numerical values for the CLD parametrization of \cite{gc20}, 
based on the SLy4 energy-density functional \cite{Chabanat_ea98_SLY4}.
In the FAC state the oi interface is located at $P=8.1\times 10^{29}$~dyn/cm$^2$ 
and the pressure at the cc boundary is $P \approx 5.20\times 10^{32}$~dyn/cm$^2$.
Assuming that the ashes are composed of $^{56}$Fe, 
the deep crustal heat release equals 
$Q^\infty\approx 0.11 \, e^{\nu_{\rm oi}/2}$~MeV/nucleon.
This heat is a sum of the heat releases at the oi interface, 
$Q_\mathrm{oi}^\infty\approx 0.07 \, e^{\nu_{\rm oi}/2}$~MeV/nucleon,
and at the cc boundary, 
$Q_\mathrm{i}^\infty\approx 0.04 \, e^{\nu_{\rm oi}/2}$~MeV/nucleon.
As shown in the Letter, for more realistic models, 
which account for the shell effects, 
$Q^\infty$ can be several times larger.


%

\end{cbunit}

\end{document}